\documentstyle[doublespacing,referee]{mn}
\newcommand{\reference}{\bibitem}
\input epsf

\def\lesssim{\mathrel{\hbox{\rlap{\hbox{\lower4pt\hbox{$\sim$}}}\hbox{$<$}}}}
\def\gtrsim{\mathrel{\hbox{\rlap{\hbox{\lower4pt\hbox{$\sim$}}}\hbox{$>$}}}}
\def\apj{ApJ}

\def\aap{A\&\hskip-1pt A}

\def\mnras{MNRAS}

\def\nat{Nature}

\newcommand{\xvec}  {\mbox{\boldmath $x$}}
\newcommand{\yvec}  {\mbox{\boldmath $y$}}

\newcommand{\uvec}  {\mbox{\boldmath $u$}}

\title[Multiple Planetary Microlensing]
      {Properties of Microlensing Light Curve\\
       Anomalies Induced by Multiple Planets}
\author[Han et al.]
{
Cheongho Han$^{1}$\thanks{cheongho@astroph.chungbuk.ac.kr}, 
Heon-Young Chang$^{2}$\thanks{hyc@ns.kias.re.kr}, 
Jin H. An$^{3}$\thanks{jinhan@astronomy.ohio-state.edu}, 
and 
Kyongae Chang$^{4}$\thanks{kchang@chongju.ac.kr}\\
$^1$Department of Physics, Chungbuk National University, 
	Chongju 361-763, Korea\\
$^2$Korea Institute for Advanced Study, 207-43 Cheongryangri-dong 
	Dongdaemun-gu, Seoul 130-012, Korea\\
$^3$Department of Astronomy, Ohio State University, Columbus, 
	OH 43210, USA\\
$^4$Department of Physics, Chongju University, Chongju 360-764, Korea
}

\begin{document}
\maketitle
\label{firstpage}
\begin{abstract}
In this paper, we show that the pattern of microlensing light curve 
anomalies induced by multiple planets are well described by the 
superposition of those of the single-planet systems where the individual 
planet-primary binary pairs act as independent lens systems.  Since the 
outer deviation regions around the planetary caustics of the individual 
planets occur in general at different locations, we find that the pattern 
of anomalies in these regions are hardly affected by the existence of 
other planet(s).  This implies that even if an event is caused by a 
multiple planetary system, a simple single-planet lensing model is good 
enough for the description of most anomalies caused by the source's 
passage of the outer deviation regions.  Detection of the anomalies 
resulting from the source trajectory passing both the outer deviation 
regions caused by more than two planets will provide a new channel of 
detecting multiple planets.
\end{abstract}


\section{Introduction}
Since the first discovery of extra-solar planets around a pulsar PSR 
1257+12 from the analysis of planetary perturbations on pulse arrival 
time by Wolszczan \& Frail (1992) and Wolszczan (1994), various groups 
have detected planets around nearby stars using high precision radial 
velocity measurements, e.g., around 51 Pegasi (Mayor \& Queloz 1995),
47 Ursae Majoris (Butler \& Marcy1996), 70 Virginis (Marcy \& Butler 1996),
and 16 Cygni B (Cochran et al.\ 1997).  Up to now, nearly 60 extra-solar 
planets are known (http://cfa-www.harvard.edu/planets).  From close 
examination of the velocity residuals to the Keplerian fits to a subsample 
of 12 planet-bearing stars, Fischer et al.\ (2001) suggested that nearly 
1/2 of the stars may have a second companion.  Currently, six systems are 
identified to have multiple planets: around $\nu$ And (Butler et al.\ 
1999), HD 83443 (Mayor et al.\ 2001), HD 168443 (Marcy et al.\ 2001a),
G7 876 (Marcy et al.\ 2001b), HD 82943 and HD 74156 
(http://exoplanets.org/mult.shtml).

Although majority of the best candidate extra-solar planets were 
discovered by the radial velocity technique, they can also be detected 
by using other techniques.  Microlensing is one of the promising 
techniques, especially to search for planets located at Galactic-scale 
distances.  Planet detections by using microlensing is possible because 
the event caused by a lens system composed of a planet (or planets) 
can exhibit noticeable anomalies in the lensing light curve when the 
source star passes closely to the lens caustics (Mao \& Paczy\'nski 1991).  
The caustic refers to the source position on which the lensing-induced 
amplification of a point-source event becomes infinity.  The lens system 
with a planet (planets) form several disconnected sets of caustics.  
Among them, one is located very close to the primary lens, ``central 
caustic'', and the other(s) is (are) located away from the primary lens, 
``planetary caustic(s)'' (Griest \& Safizadeh 1998).  The planet-induced 
anomalies last only a few hours to days, and thus it is difficult to 
detect them from the survey-type lensing experiments (MACHO: Alcock et 
al.\ 1993; OGLE: Udalski et al.\ 1993; EROS: Aubourg et al.\ 1993; 
MOA: Abe et al.\ 1997).  However, with sufficiently frequent and accurate 
monitoring from follow-up observations of ongoing events alerted by the 
survey teams, one can detect the anomalies and determine the planet 
parameters of the mass ratio and the instantaneous projected separation 
between the planet and the primary lens.  Currently, several groups 
are actually carrying out such observations: MPS (Rhie et al.\ 1999), 
PLANET (Albrow wt al.\ 1998), and MOA (Abe et al.\ 1997).

Along with the observational efforts, planetary microlensing has also 
been the field of intensive theoretical studies.  These include 
phenomenological studies of the lensing behaviors (Wambsganss 1997; 
Dominik 1999; Safizadeh et al.\ 1999), estimation of the detection 
probability and efficiency (Gould \& Loeb 1992; Bolatto \& Falco 1994; 
Bennett \& Rhie 1996; Peale 1997, 2001; Gaudi \& Sackett 2000), 
development of the observational strategies for optimal planet 
detections (Griest \& Safizadeh 1998; Han \& Kim 2001), and various 
problems and solutions in determining planet parameters (Gaudi \& Gould 
1997; Gaudi 1998; Griest \& Safizadeh 1998).  However, most of these 
studies were focused on lens systems composed of only a single planet, 
despite the fact that our solar system is composed of multiple planets.  
Gaudi, Naber \& Sackett (1998) demonstrated that if extra-solar systems 
are composed of planets having orbital separations comparable to those 
of Jupiter and Saturn of our solar system, the joint probability of 
two planets having projected separations within the lensing zone of 
$\sim 0.6$ -- 1.6 of the Einstein ring radius is substantial, invoking 
the necessity of studies about the lensing properties of multiple 
planetary systems.

In this paper, we investigate the properties of lensing light curve 
anomalies induced by multiple planets.  Multiple planetary microlensing 
was first explored by Gaudi et al.\ (1998).  However, since they were 
interested only in the anomalies occurring near the peaks of the light 
curves of high amplification events, for which the planet detection 
efficiencies are high (Griest \& Safizadeh 98), they investigated 
anomaly patterns only in the region around central caustics.  In this 
work, we extend the investigation in wider areas including the outer 
deviation regions around planetary caustics.

The paper is organized in the following way.  We begin with a brief 
review of the basics of multiple planetary lensing in \S\ 2.  In \S\ 3,
we present the anomaly patterns of an example lens system and the 
resulting light curves to illustrate the effect of multiple planets.  
We discuss about the implications of the new findings in \S\ 4.  A 
brief comment and conclusion are found in \S\ 5.

\section{Basics of Multiple Planetary Lensing}

A lens system composed of multiple planets is described by the formalism 
of multiple lensing with very low mass-ratio companions.  If a source star 
is lensed by $N$ point-mass lenses, the locations of the resulting 
images are obtained by solving the mapping equation (lens equation), which 
is expressed in complex notations by
\begin{equation}
\zeta=z+\sum_j^N {m_j\over \bar{z}_j - \bar{z}},
\end{equation}
where $m_j$'s represent the mass fractions of the individual lenses such
that $\sum_j^N m_j=1$, $z_j$'s are the positions of the lenses, $\zeta=
\xi+i\eta$ and $z=x+iy$ are the positions of the source and images, and 
$\bar{z}$ denotes the complex conjugate of $z$ (Witt 1990).  We note 
that all these lengths are normalized by the angular Einstein ring radius, 
which is related to the total mass, $M$, and the geometry of the lens 
system by
\begin{equation}
\theta_{\rm E}=\sqrt{4GM\over c^2}
\left( {1\over D_{\rm ol}}-{1\over D_{\rm os}} \right)^{1/2},
\end{equation}
where $D_{\rm ol}$ and $D_{\rm os}$ represent the distances to the lens 
and source from the observer, respectively.  The angular separations 
between the microlensed images are of the order of several milli-arcsecs, 
which are too small to be resolved.  However, the lensing event can be 
identified by the change of the source star flux.  The amplifications of 
the individual images are given by the inverse of the determinant of the 
Jacobian of the lens equation evaluated at the image position, i.e.
\begin{equation}
A_i=\left( \frac{1}{|{\rm det}\ J|} \right)_{z=z_i};\ \ 
    {\rm det}\ J=1-{\partial \zeta \over \partial \bar{z}}
     {\overline{\partial \zeta} \over \partial \bar{z}}.
\end{equation}
Then the total amplification is obtained by the sum of those of the 
individual images, i.e.\ $A=\sum_{i}^{N_I} A_i$, where $N_I$ represents 
the number of images.  The caustic corresponds to the source position on 
which the determinant becomes zero, i.e.\ $|{\rm det}\ J|=0$.  The lens 
equation describes a mapping from the lens plane onto the source plane.  
Then, to find image positions $(x,y)$ for a given source position 
$(\xi,\eta)$, the equation should be inverted.

For a single lens system ($N=1$), the lens equation can be inverted and 
thus is algebraically solvable, yielding two solutions (and thus the 
same number of images).  The total amplification is expressed in a 
simple form of
\begin{equation}
A = {u^2+2\over u \sqrt{u^2+4}},
\end{equation}
where $\uvec$ is the dimensionless lens-source separation vector normalized 
by $\theta_{\rm E}$.  The separation vector is related to the single lensing 
parameters by 
\begin{equation}
\uvec\ =\ 
\left({t-t_{0}\over t_{\rm E}}\right)\ \hat{\xvec}\ +
\ \beta\ \hat{\yvec}\ ,
\end{equation}
where $t_{\rm E}$ represents the time required for the source to transit 
$\theta_{\rm E}$ (Einstein time scale), $\beta$ is the closest lens-source 
separation in units of $\theta_{\rm E}$ (impact parameter), and $t_0$ is 
the time at that moment. The unit vectors $\hat{\xvec}$ and $\hat{\yvec}$ 
are parallel with and normal to the direction of the relative lens-source 
transverse motion.

For a multiple lens system ($N\geq 2$), on the other hand, the lens 
equation is non-linear and thus cannot be inverted.  However, it can be 
expressed as a polynomial in $z$, and the image positions are obtained 
by numerically solving the polynomial.  For a $N$ point-mass lens system, 
the lens equation is equivalent to a $(N^2+1)$-order polynomial in $z$ 
and there are a maximum $N^2+1$ and a minimum $N+1$ images and the number 
of images changes by a multiple of two as the source crosses a caustic.  
Due to the non-linear nature of the lens mapping for a multiple lens 
system, the lensing behavior of events caused by a multiple lens is 
{\it not} given by a simple superposition of the individual single lens 
events.

Unless the planet has a separation very close to $\theta_{\rm E}$, a 
single-planet lens system forms a single central caustic and one or two 
planetary caustics depending on the projected separation between the 
primary and the planet.  Accordingly, there exist two different types of 
planet-induced anomalies; one affected by the planetary caustic(s) 
(``type I anomaly'') and the other affected by the central caustic 
(``type II anomaly'') (Han \& Kim 2001).  Compared to the frequency of 
type I anomalies, type II anomalies occur with a relatively low frequency 
due to the smaller size of the central caustic compared to that of the 
planetary caustic(s).  However, the efficiency of detecting type II 
anomalies can be high because intensive monitoring is possible due to 
the known type of target events (i.e., high amplification events) and 
the predictable time of anomalies (i.e., near the peak of light curves).

As the number of lenses increases, solving the lens equation becomes 
very cumbersome.  One commonly practiced approach to obtain amplification 
patterns of these lens systems is the inverse ray-shooting technique 
(Schneider \& Weiss 1986; Kayser, Refsdal \& Stabell 1986; Wambsganss 1997).  
In this method, a large number of light rays are shoot backwards from 
the observer plane through the lens plane, and then collected (binned) 
in the source plane.  Then the amplification pattern is obtained by 
the ratio of the surface brightness (i.e., the number of rays per unit 
area) on the source plane to that on the observer plane and the light 
curve from a particular source trajectory corresponds to the 
one-dimensional cut through the constructed amplification pattern.  
This method has an advantage of enabling one to obtain amplification 
patterns for an arbitrary number of lenses, but has a disadvantage of 
requiring large computation time to obtain smooth amplification patterns.

In our analysis of multiple-planet lensing properties, we test lens systems 
composed of two Jovian-mass planets (and thus $N=3$).  This is because 
if extra-solar systems have planets with orbital separations and mass 
ratios similar to those of our system, other planets except the two 
Jovian planets (Jupiter and Saturn) will have mass ratios too small 
to affect the amplification pattern.  In addition, the probability of more 
than three planets being simultaneously located in the lensing zone will 
be very small.  Besides, this allows us to directly solve the lens 
equation rather than using the more time-consuming ray-shooting method.

\section{Properties of Anomalies}

For the investigation of the properties of lensing light curve anomalies 
induced by multiple planets, we construct maps of amplification excesses. 
The amplification excess is defined by
\begin{equation}
\epsilon_{\rm tri} = {A_{\rm tri}-A_0 \over A_0},
\end{equation}
where $A_{\rm tri}$ and $A_0$ represent the amplifications with and 
without planets.  With the map, one can obtain an overview of the 
anomaly patterns not testing all light curves resulting from a large 
number of source trajectories.

In the upper panel of Figure 1, we present the constructed excess map of 
an example lens system.  The lens system has two planets with mass ratios 
of $q_1=0.003$ and $q_2=0.001$, which respectively correspond to those of 
Jupiter- and Saturn-mass planets around a $\sim 0.3\ M_\odot$ star.  The 
individual planets have projected separations (normalized by $\theta_{\rm E}$) 
of $b_1=1.3$ and $b_2=0.85$ and the orientation angle between the position 
vectors from the primary lens to the individual planets is $\phi=150^\circ$.  
In the map, the levels of excesses are represented by contours, which are 
drawn at the levels of $\epsilon=-10\%$, $-5\%$, 5\%, and 10\% and the 
regions of positive excess are distinguished by grey scales.  To show 
the detailed excess patterns in the region around the central caustic, 
the map of the region is expanded and presented separately in Figure 2.

From the investigation of the map, one may notice an interesting point 
that the locations of the outer caustics are very similar to those of the
planetary caustics of the individual planet-primary binary pairs.  To show 
this similarity more closely, in the lower panel of Fig.\ 1 we present the 
excess map constructed by superposing the excesses of the two single-planet 
lens systems where the individual planet-primary pairs act as independent 
lens systems (binary superposition), i.e.
\begin{equation}
\epsilon_{\rm bs} = \sum_{i=1}^2 {A_{{\rm b},i}-A_0\over A_0},
\end{equation}
where $A_{{\rm b},i}$'s represent the binary-lensing amplifications of the 
individual planet-primary pairs.  In Figure 3, we also present the blowup 
of the central region of the map.  From the comparison of the excess 
patterns in the outer deviation regions, one finds that not only the 
locations of the outer caustics but also the excess patterns around them 
are very similar each other.  This can be seen also from the comparison 
of the light curves of events caused by the triple lens system (solid 
curves in the left panels of Figure 5) to those obtained by the binary 
superposition (dotted curves in the same panels of the same figure), 
which are obtained by
\begin{equation}
A_{\rm bs}=A_0 (1+\epsilon_{\rm bs})
 =  \sum_{i=1}^2 A_{{\rm b},i} - A_0.
\end{equation}
From the comparison of the maps in Fig.\ 2 and 3, one finds that the 
excess patterns in the region around the central caustic are also 
similar each other, although the similarities are not as exact as those 
in the outer deviation region.\footnote{For example, compared to the 
two disconnected sets of closed shape for the caustics constructed by 
the binary superposition, the triple lens system forms a single set of 
nested caustics.}

To more systematically investigate how accurately the multiple-planet 
lensing properties can be represented as binary superposition, we compute 
the fractional deviation of the amplification obtained by the binary 
superposition from that of the exact triple lens system by
\begin{equation}
\delta = {A_{\rm tri}-A_{\rm bs}\over A_{\rm tri}}.
\end{equation}
In Figure 4, we present the contour map of $\delta$ for the same lens 
system whose excess map is presented in Fig.\ 1.  We note that in the 
map the contours are drawn at the levels of $\delta=-1\%$, $-0.5\%$, 
0.5\%, and 1\%, which are much smaller than those of the contours in 
the map of $\epsilon$.  In the right panels of Fig.\ 5, we also present 
the deviation curves of $\delta$ for the events resulting from the 
source trajectories marked in Fig.\ 1.  From the map and deviation 
curves, one finds that the differences occur only in small localized regions 
very close to the caustics.  The spikes in the deviation curves are 
caused by the shift of the caustics.  However, we note that the shift 
is very small ($\sim 10^{-3} \theta_{\rm E}$), corresponding to the 
time shift of approximately half an hour for a typical Galactic bulge 
event with $t_{\rm E}\sim 20$ -- 30 days.  Considering the photometric 
precision ($\gtrsim 1\%$) and monitoring frequency ($\sim 3$ times/night) 
of the current microlensing follow-up observations, it will be difficult 
to notice the caustic shifts.  In addition, the modified light curve 
can be fit by a light curve with very slightly changed lensing parameters.
Therefore, the binary superposition will be a good approximation for 
the description of the example multiple-planet system.

Then, to what companion mass ratios is the approximation of binary 
superposition valid for the description of the lensing behavior of 
multiple-companion systems?  To answer this question, we compute the 
fraction of the area within the Einstein ring where the binary superposition 
approximation deviates more than a threshold value $\delta_{\rm th}$ for 
lens systems with various combinations of mass ratios, $f_{\rm dev}
(\delta_{\rm th})$.  In Figure 6, we present the contours of constant 
$f_{\rm dev}$ in the parameter space of ${\rm log}\ q_1$ and 
${\rm log}\ q_2$.  We note that the presented $f_{\rm dev}$ are the 
values averaged for lens systems with planets having relative orientation 
angles in the range $0\leq \phi\leq 2\pi$ and located in the lensing zone,
i.e.\ $0.6 \lesssim b_i \lesssim 1.6$.  The upper and lower panels 
represent the maps constructed with the threshold deviations of 
$\delta_{\rm th}=0.5\%$ and $1\%$, respectively.  From the maps, one 
finds that most lens systems with heavier companion mass ratios of 
$q_1\lesssim 10^{-2}$ have $f_{\rm dev} \leq 1\%$, implying that the 
lensing behaviors of multiple-planet systems are well approximated by 
the binary superposition.  We note, however, that if $q_1\gtrsim 0.05$, 
the fraction becomes considerable [$f_{\rm dev}
(\delta_{\rm th}=0.5\%)\gtrsim 8\%$ and $f_{\rm dev}(\delta_{\rm th}=1\%)
\gtrsim 4\%$], implying that the binary superposition approximation 
will not be appropriate to describe the lensing behavior of systems 
composed of very heavy planets 
(with masses $\gtrsim 10\ M_{\rm J}$)\footnote{For HD 168443, which 
was identified to have multiple planets by the radial velocity technique, 
one planet member has a mass of $\gtrsim 17\ M_{\rm J}$} and 
brown-dwarf-mass companions.

\section{Implications}

In the previous section, we show that the patterns of amplification 
excesses of a lens system composed of multiple planets can be well 
approximated by the superposition of excesses induced by the planets 
of the individual planet-primary binary pairs.  This finding has the 
following implications in describing the lensing behaviors of events 
caused by multiple planetary systems and determining their planet 
parameters.

The first implication of our finding is that even if an event is caused 
by a multiple-planet lens system, a simple single-planet lensing model 
will be good enough for the description of most type I anomalies.  This 
is because the excess pattern in the outer deviation region caused by 
one of the planets is hardly affected by the other planet component(s).  
Complication might occur if the outer deviation regions induced by the 
individual planets occur at a similar place.  However, this case is 
rare since the outer excess region induced by a planet-mass companion 
occupies a small area.

However, the situation becomes complicated for the type II anomalies.  This 
is because the central deviation regions of the individual planets always 
occur in the same region regardless of their orientations.  As a result, 
the pattern of anomalies induced by one planet in this region can be 
significantly affected by the existence of other planet(s), as pointed out 
by Gaudi et al.\ (1998).  Since one should apply complex multiple-planet 
lensing models for the description of anomalies occurred in this region, 
accurate determinations of the planet parameters from the observed light 
curves will be difficult, as pointed out also by Gaudi et al.\ (1998).

\section{Conclusion}

We have investigated the properties of anomalies in lensing light curves
induced by multiple planets.  From the analysis of the maps of excess 
amplification and resulting light curves, we find that the excess 
patterns of a multiple-planet lens system are well described by the 
superposition of the excesses of the single-planet systems where the 
individual planet-primary pairs act as independent lens systems.  
This finding has an important implication that a simple single-planet 
lensing model is good enough for the description of most type I 
anomalies even if an event is caused by a lens system composed of 
multiple planets.

Another interesting point to be mentioned is that detection of the 
anomalies resulting from the source trajectory passing both the outer 
regions of deviation induced by more than two planets (e.g., the 
trajectory designated by a number `3' in Fig.\ 1) will provide a new 
channel of detecting multiple planets.  The previously suggested method 
of detecting multiple planets by analyzing anomalies occurred by the 
source's passage of the central caustic region (such as the trajectory 
`4' in Fig.\ 1) has significant limitations in identifying the existence 
of multiple planets and determining their parameters due to the degeneracy 
of the resulting light curves from those of single-planet lensing and 
the complexity of multiple-planet lensing models.  On the other hand, 
if a multiple-planet lens system is detected through the new channel, 
one can better determine the planet parameters because the individual 
anomalies are well approximated by relatively simple single-planet 
lensing models.  We defer a more systematic analysis of this method 
and estimation of multiple planet detection probabilities for future 
work.

We thank to B.\ S.\ Gaudi for making helpful comments on multiple planetary
lensing.  This work was supported by a grant (2000-015-DP0449) from the 
Korea Research Foundation (KRF).

{}

\clearpage

\begin{figure*}
\epsfysize=17cm
\centerline{\epsfbox{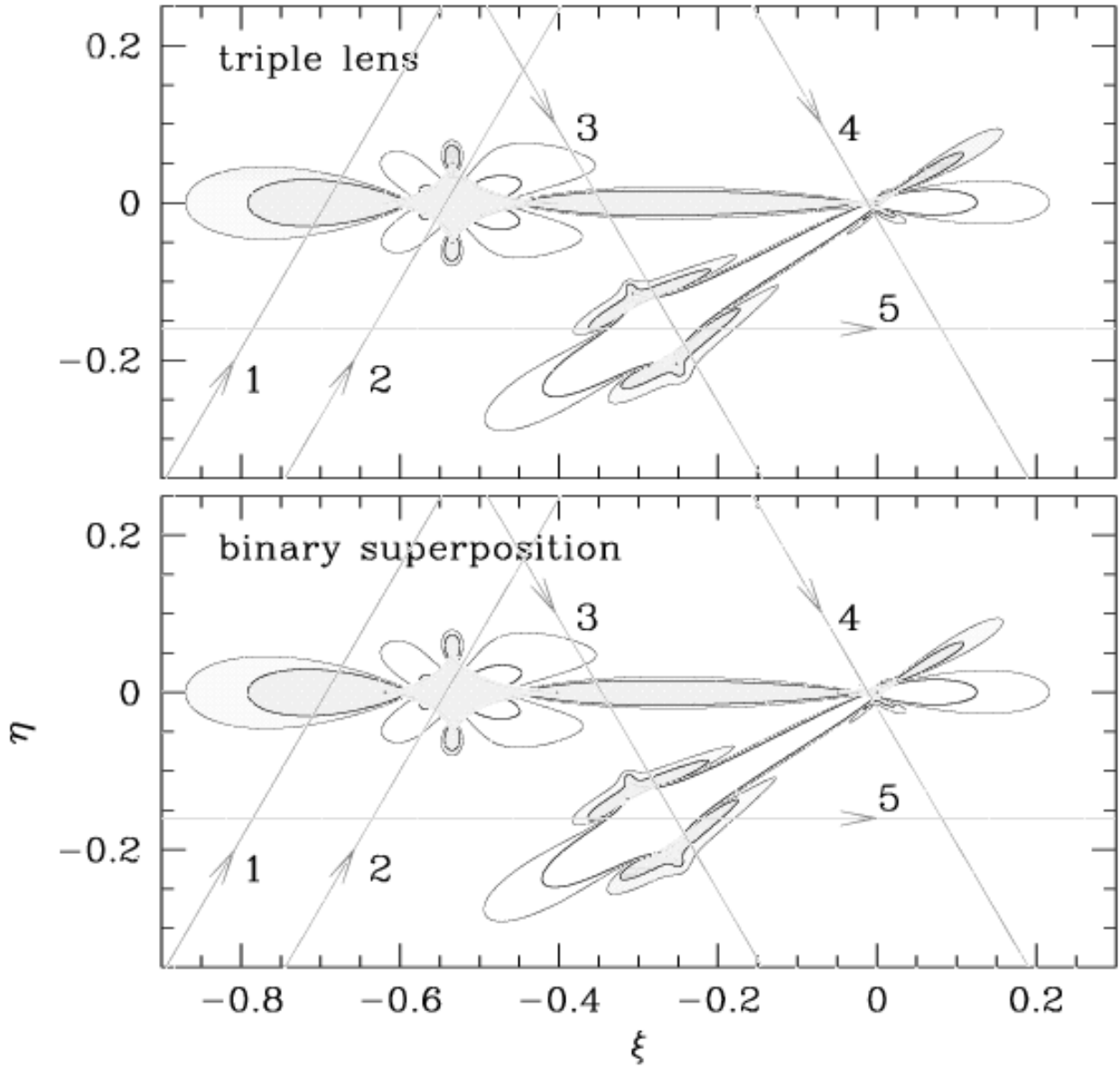}}
\caption{
Upper panel: The contour map of amplification excesses of a lens system
composed of multiple planets.  The lens system has two planets with the
ratios of mass to that of the primary of $q_1=0.003$ and $q_2=0.001$ and 
the projected separations from the primary of $b_1=1.3$ and $b_2=0.85$, 
respectively.  The locations are set so that the primary is at the center.  
The heavier planet is located on the $\xi$ axis on the left side of the 
primary and the orientation angle between the position vectors to the 
individual planets (with respect to the primary) is $150^\circ$.  The 
contours are drawn at the levels of $\epsilon = -10\%$, $-5\%$, 5\% and 
10\% and the regions of positive excesses are distinguished by grey scales.  
Lower panel: The excess map constructed by the superposition of excesses 
of the single-planet lens systems where the individual planet-primary 
pairs acts as independent lens systems.  For both maps, the figures drawn 
by thick solid line represent the caustics and the straight lines with 
arrows are the source trajectories of events whose resulting light and 
excess curves are presented in Fig.\ 5.
}
\end{figure*}

\begin{figure*}
\epsfysize=17cm
\centerline{\epsfbox{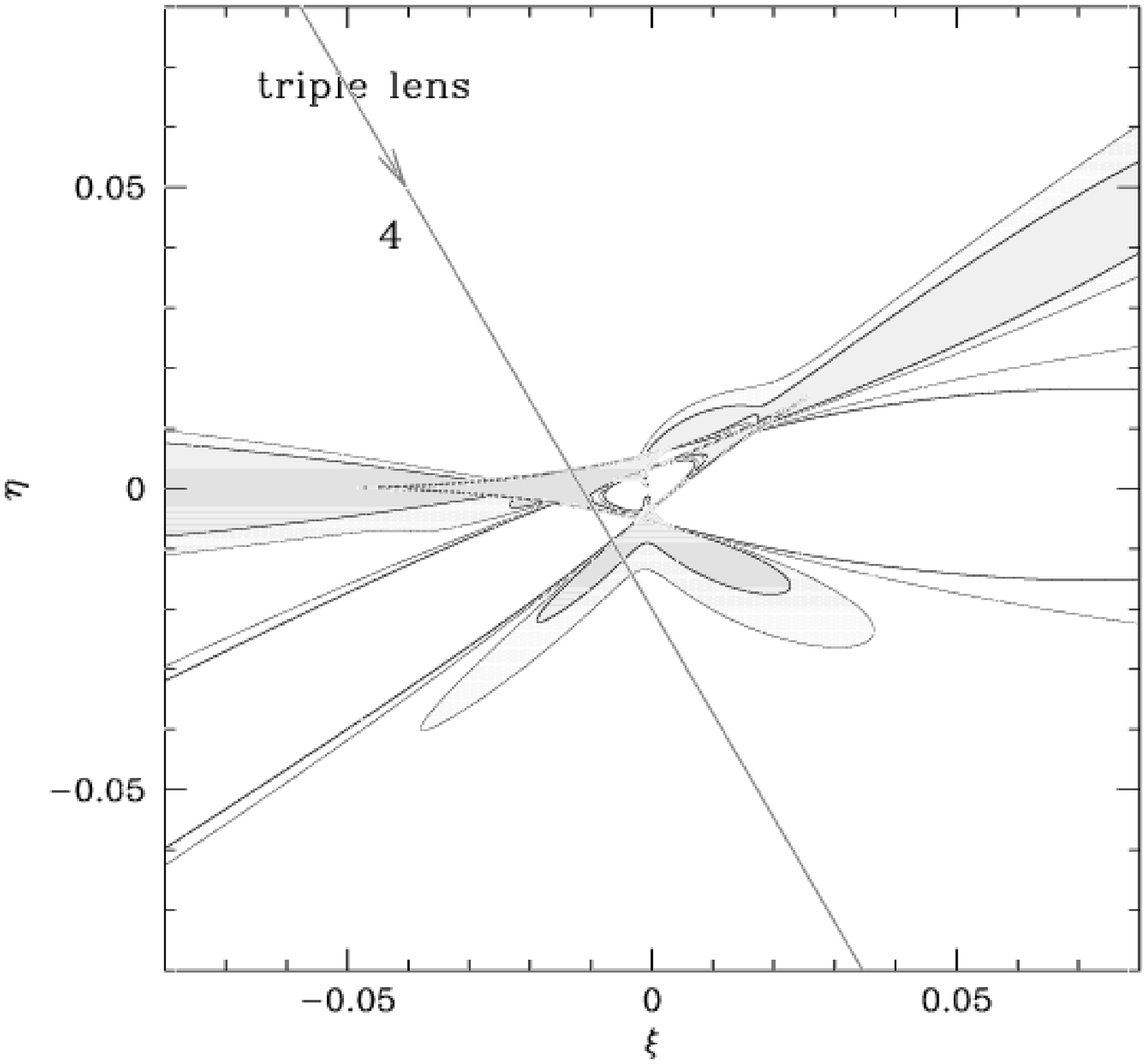}}
\caption{
The blowup of the central caustic region of the amplification excess map 
presented in the upper panel of Fig.\ 1.
}
\end{figure*}

\begin{figure*}
\epsfysize=17cm
\centerline{\epsfbox{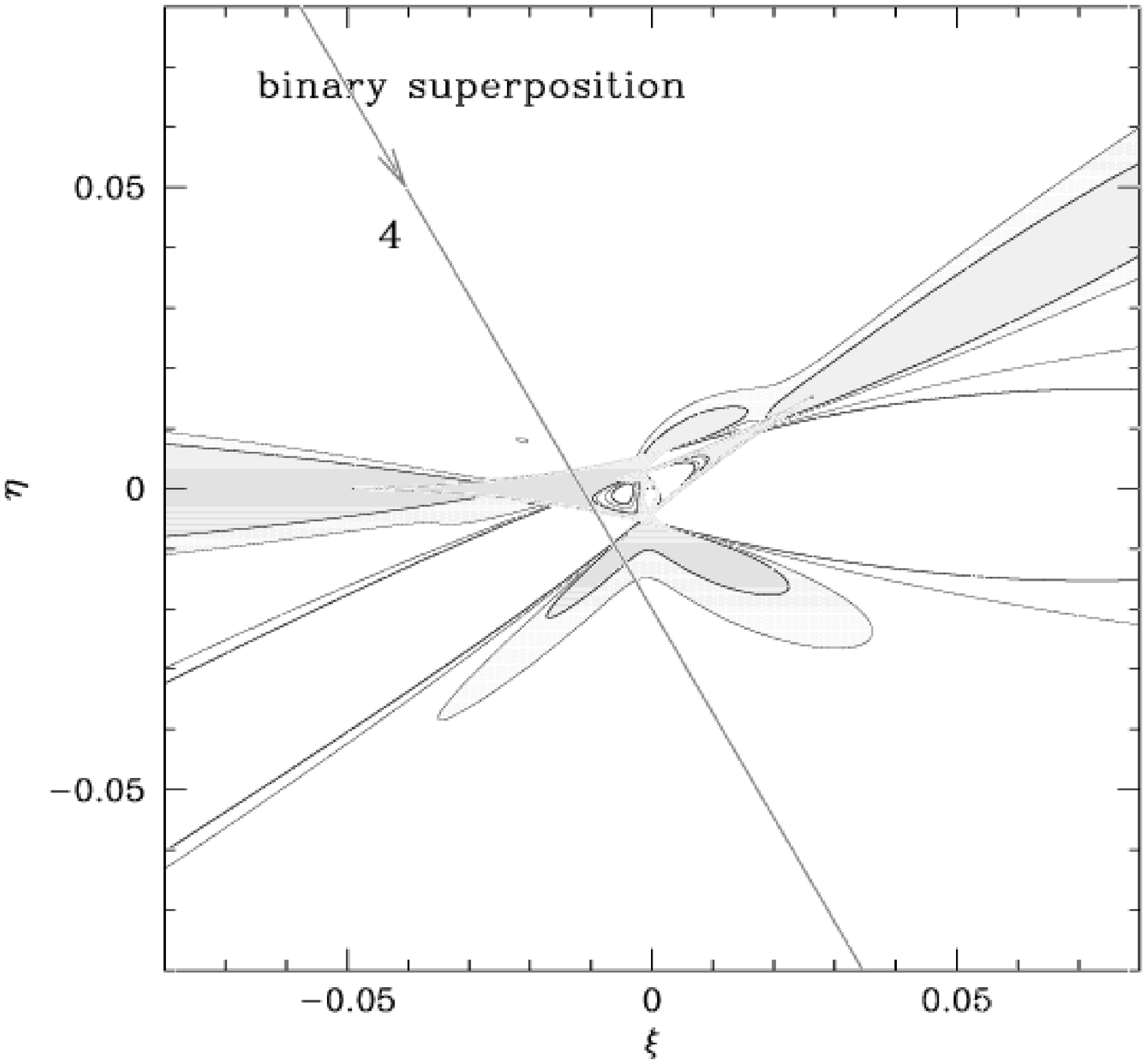}}
\caption{
The blowup of the central caustic region of the amplification excess map 
presented in the lower panel of Fig.\ 1.
}
\end{figure*}

\begin{figure*}
\epsfysize=17cm
\centerline{\epsfbox{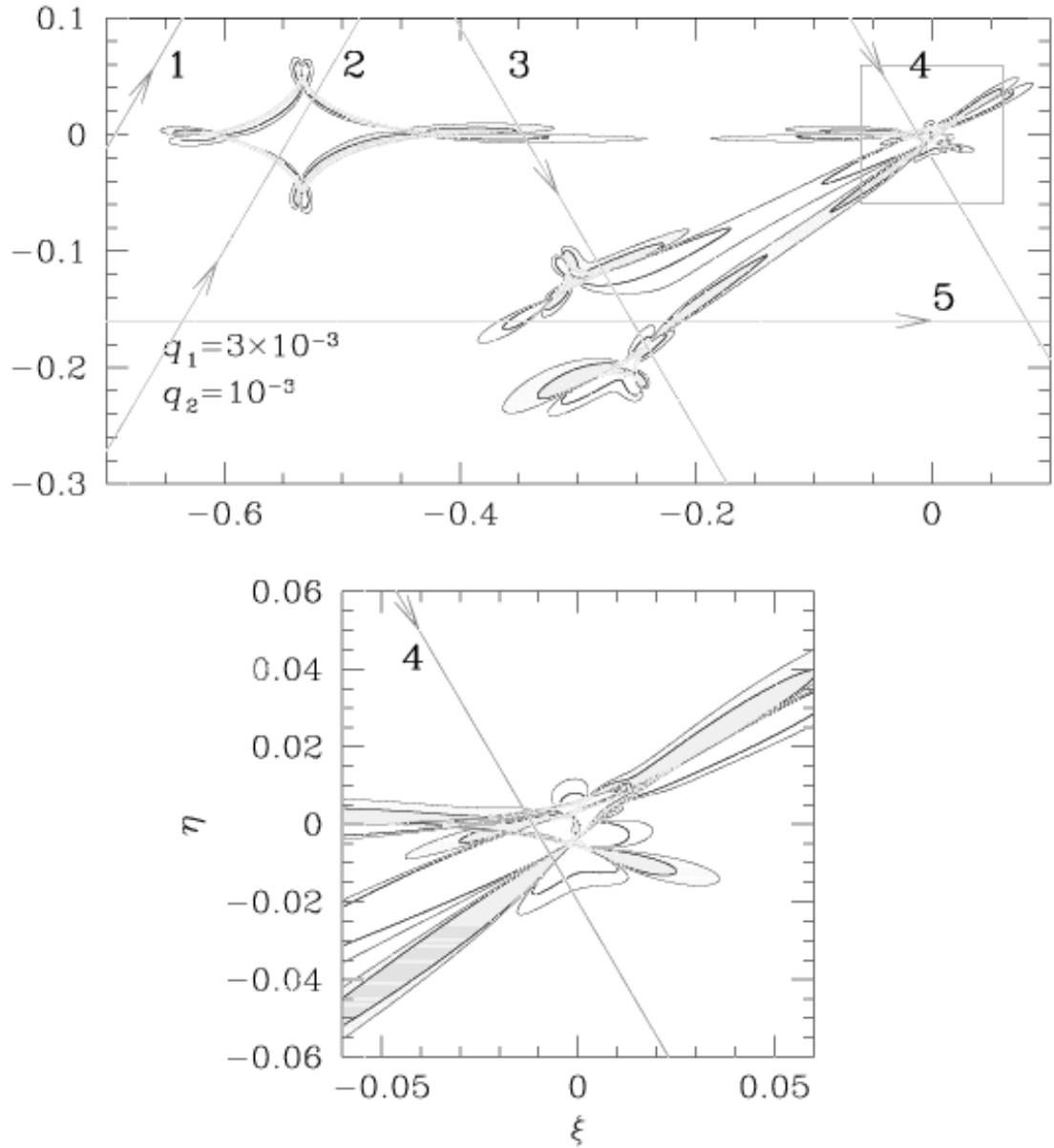}}
\caption{
The contour map of the fractional deviation of the amplification obtained 
by the binary superposition approximation from that of the exact triple 
lens system.  The map is for the same lens system whose excess map is 
presented in Fig.\ 1.  The contours are drawn at the levels of 
$\delta=-1\%$, $-0.5\%$, 0.5\%, and 1\% and the regions of positive 
$\delta$ is distinguished by grey scales.  The lower panel is the blowup 
of the map in the region around the central caustic.
}
\end{figure*}

\begin{figure*}
\epsfysize=17cm
\centerline{\epsfbox{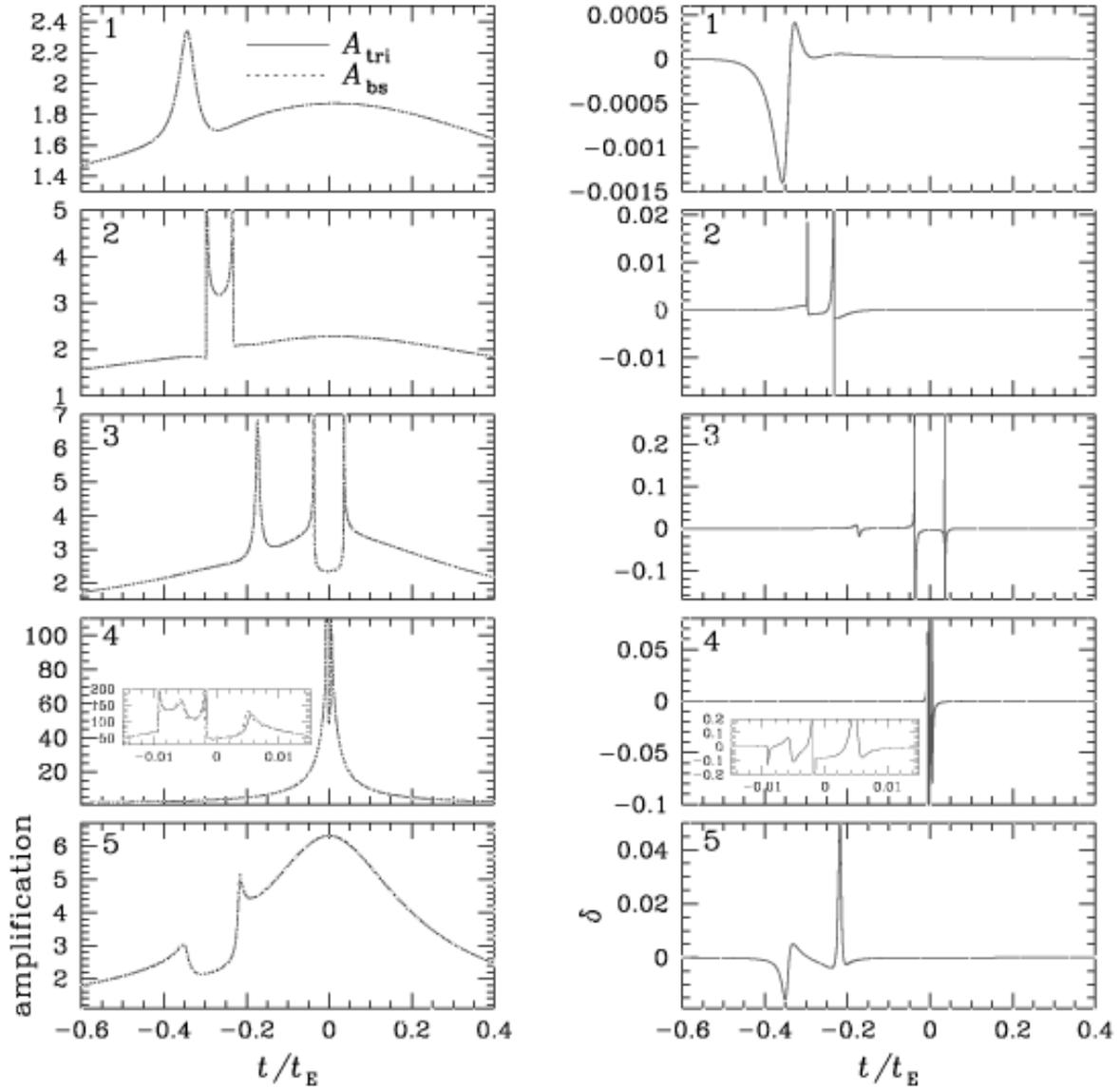}}
\caption{
Left panels: 
The light curves of the events resulting from the source trajectories
marked in Fig.\ 1. The number in each panel corresponds to the trajectory 
number.
Right panels: 
The curves of the fractional deviations of the amplification obtained by 
the binary superposition approximation from that of the exact triple lens 
system.
}
\end{figure*}

\begin{figure*}
\epsfysize=17cm
\centerline{\epsfbox{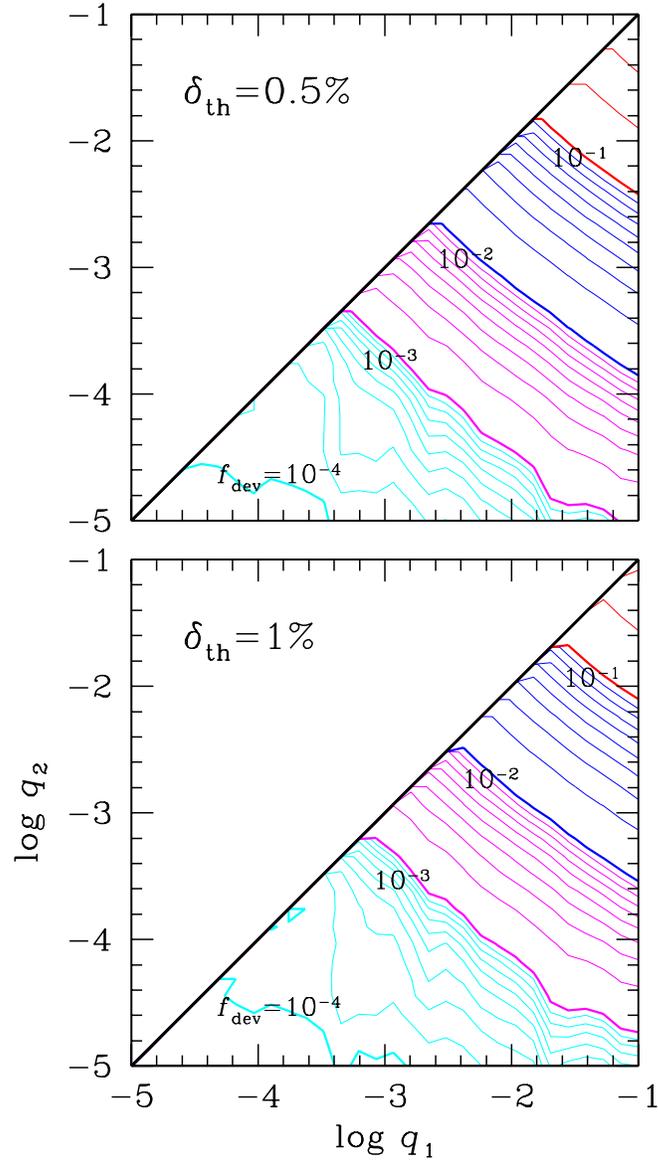}}
\caption{
The contours of constant $f_{\rm dev}$, which represents the fraction
of the area within the Einstein ring where the binary superposition
approximation deviates from the exact lensing amplification more than
a threshold value $\delta_{\rm th}$ for the lens systems composed of
two planets with mass ratios $q_1$ and $q_2$.  The presented $f_{\rm dev}$
are the values averaged for lens systems with planets having relative
orientation angle in the range $0\leq \phi \leq 2\pi$ anf located in
the lensing zone, i.e.\ $0.6 \lesssim b_i \lesssim 1.6$.  The upper and 
lower panels are the maps constructed with the threshold deviations of 
$\delta_{\rm th}=0.5\%$ and $1\%$, respectively.
}
\end{figure*}

\end{document}